\documentclass[conference]{IEEEtran}
\IEEEoverridecommandlockouts
\usepackage{cite}
\usepackage{amsmath,amssymb,amsfonts}
\usepackage{algorithmic}
\usepackage{graphicx}
\usepackage{textcomp}
\usepackage{xcolor}

\usepackage{xspace}
\usepackage{amsmath,color,soulutf8,longtable,colortbl,setspace,xspace,url,pdflscape,cite}
\usepackage{supertabular,booktabs}
\usepackage{pgf}
\usepackage{multirow}  

\def\BibTeX{{\rm B\kern-.05em{\sc i\kern-.025em b}\kern-.08em
    T\kern-.1667em\lower.7ex\hbox{E}\kern-.125emX}}
\begin{document}

\newcommand{\ie}{\emph{i.e.,}\xspace}
\newcommand{\eg}{\emph{e.g.,}\xspace}
\newcommand{\etc}{\emph{etc.}\xspace}
\newcommand{\etal}{\emph{et al.}\xspace}

\newcommand{\bugname}[1]{\texttt{#1}}

\newcommand{\framework}{\texttt{ReproduceML}}

\newcommand{\corrected}{corrected}
\newcommand{\buggy}{buggy}

\newcommand{\IAC}{Industry-Academia Collaboration}
\newcommand{\pytorch}{PyTorch}
\newcommand{\tensorflow}{TensorFlow}

\newcommand{\ai}{AI}
\newcommand{\Ml}{ML}
\newcommand{\ML}{ML}
\newcommand{\ml}{ML}
\newcommand{\DL}{DL}
\newcommand{\CNN}{CNN}

\newcommand{\dl}{DL}
\newcommand{\gpu}{GPU}
\newcommand{\GPU}{GPU}
\newcommand{\CPU}{CPU}

\newcommand{\lstm}{LSTM}
\newcommand{\rnn}{RNN}
\newcommand{\dnn}{DNN}
\newcommand{\dnns}{DNNs}
\newcommand{\VCS}{VCS}
\newcommand{\hypp}{hyperparameter}
\newcommand{\ndif}{NDIF}
\newcommand{\pct}{\%}
\newcommand{\cfvars}{confounding variables}

\newcommand{\programmingLanguage}[1]{#1}
\newcommand{\Python}{\programmingLanguage{Python}}
\newcommand{\R}{\programmingLanguage{R}}

\newcommand{\vcs}{VCS}
\newcommand{\vcss}{VCSs}
\newcommand{\API}{API}
\newcommand{\PCA}{PCA}
\newcommand{\SVD}{SVD}

\newcommand{\extf}[1]{\textbf{#1}}
\newcommand{\extlib}[1]{\textbf{#1}}
\newcommand{\deepdiva}{\extf{DeepDIVA}\cite{Alberti2019}}

\newcommand\sideeffects{side effects}
\newcommand\MergeR{Merge Request}
\newcommand\PullR{Pull Request}

\newcommand{\mbuggy}{\textbf{buggy}}
\newcommand{\mcorrected}{\textbf{corrected}}
\newcommand{\mversions}{\mbuggy~and \mcorrected}
\newcommand{\keypoint}[1]{\textbf{#1}}
\newcommand{\git}{\texttt{Git}}
\newcommand{\CUDA}{\texttt{CUDA}}
\newcommand{\run}{run}
\newcommand{\wmw}{Wilcoxon-Mann-Whitney}
\newcommand{\Docker}{\emph{Docker}}
\newcommand{\server}{\emph{server}}
\newcommand{\client}{\emph{client}}
\newcommand{\trainingIpcEvaluation}{training-ipc-evaluation}
\newcommand{\mstep}[2]{\textbf{#1}: #2\\}
\newcommand{\bugid}[1]{\texttt{#1}}

\newcommand{\version}[1]{\text{#1}}
\newcommand{\nBugsCollected}{737}
\newcommand{\nBugsLabelled}{439}
\newcommand{\nBugsUsedForStudy}{115}
\newcommand{\nBugsBuilt}{49}
\newcommand{\nBugsResults}{18}
\newcommand{\nBugsExcluded}{3}

\newcommand{\Foutse}[1]{\textcolor{blue}{{\it [Foutse: #1]}}}
\newcommand{\Amin}[1]{\textcolor{red}{{\it [Amin says: #1]}}}

\title{The Challenge of Reproducible ML: An Empirical Study on The Impact of Bugs\\

}


\author{\IEEEauthorblockN{Emilio Rivera-Landos, Foutse Khomh, Amin Nikanjam}
\IEEEauthorblockA{SWAT Lab., Polytechnique Montreal\\
Quebec, Canada\\
Email: \{emilio.rivera, foutse.khomh, amin.nikanjam\}@polymtl.ca}}


\maketitle

\begin{abstract}
Reproducibility is a crucial requirement in scientific research. When results of research studies and scientific papers have been found difficult or impossible to reproduce, we face a challenge which is called reproducibility crisis. Although the demand for reproducibility in Machine Learning (ML) is acknowledged in the literature, a main barrier is inherent non-determinism in ML training and inference. In this paper, we establish the fundamental factors that cause non-determinism in ML systems. A framework, \framework, is then introduced for deterministic evaluation of ML experiments in a real, controlled environment. \framework~allows researchers to investigate software configuration effects on ML training and inference. Using \framework, we run a case study: investigation of the impact of bugs inside ML libraries on performance of ML experiments. This study attempts to quantify the impact that the occurrence of bugs in a popular ML framework, PyTorch, has on the performance of trained models. To do so, a comprehensive methodology is proposed to collect buggy versions of ML libraries and run deterministic ML experiments using \framework. Our initial finding is that there is no evidence based on our limited dataset to show that bugs which occurred in PyTorch do affect the performance of trained models. The proposed methodology as well as \framework~ can be employed for further research on non-determinism and bugs.
\end{abstract}


\section{Introduction}\label{sec:intro}
The reproducibility of scientific results is an important topic not only in Machine Learning (ML) but in research in general. It has happened many times that one reads scientific publications showing interesting results but finds them difficult to reproduce. This is called the reproducibility crisis, a methodological phenomenon where results of research studies and scientific papers have been found to be difficult or impossible to reproduce. The reproducibility crisis in ML is acknowledged in the literature \cite{gardner2018enabling,olorisade2017reproducibility} and many propositions for research guidelines have emerged by showcasing reproducible research \cite{gardner2018enabling}, creating guidelines for research formulation \cite{doshi2017towards}, guidelines for reproducible research \cite{Stodden2013,Crook2013reproNeuro,Piccolo2016,Davison2012repro} or even by modelling research \cite{Stodden2020,vanschoren2012experiment} and its workflows \cite{Khan2019,CohenBoulakia2017}. Pham \etal surveyed the literature to understand the state of reproducibility and repeatability in Software Engineering and Artificial Intelligence (\ai), and found that only 19.5\% of papers tried to use several identical runs when reporting their results \cite{Pham}. Similarly, a recent analysis of code inclusion in Deep Learning (\DL) papers found that 25.8\% of them include code \cite{liu2020replicability}. While code inclusion in a research paper goes one step further for reproducibility, it needs to be usable. A study by Gunderson \cite{gundersen2018state} introduces degrees of reproducibility, and finds that most papers in \ai~are not fully reproducible. A similar study for \ml~applied in medicine shows a 9\% reproducibility \cite{felix2020variability}. Other researchers have also called for better reproducibility in \ml \cite{Raeder2010,olorisade2017reproducibility}~and \dl~\cite{jean2019issues, khetarpal2018re}.

Reproducibility of the results is directly impacted by the underlying causes of variance: \textit{Non-Determinism Introducing Factors} (\ndif). In the context of ML, we define \ndif~as any change in the ML lifecycle that directly or indirectly affects the training or inference results due to non-deterministic behaviours. In other words, NDIF are factors that are important to consider for reproducing an experiment. Borrowed from Pham \etal \cite{Pham} and adapted with findings of Crane \cite{Crane2018} and Guo \etal \cite{Guo2019}, we present seven main NDIF that cause non-determinism in \ml: random seed, model definition, software versions, threading model, runtime, hardware and data. Each of these NDIF finds their non-determinism in three root causes: randomness, versioning, and floating point operations. In the following, we briefly review these factors.

\begin{figure*}[ht]
    \centering
    \includegraphics[width=0.9\textwidth]{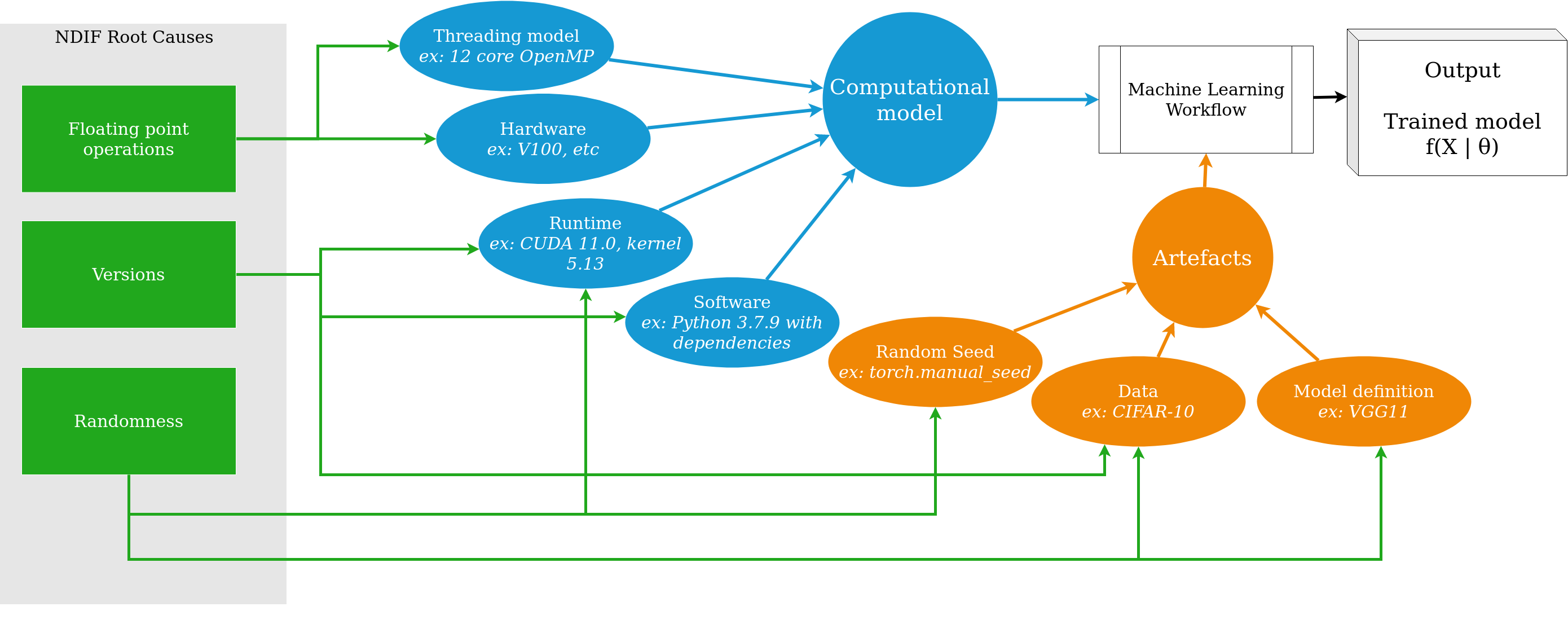}
    \caption{Non-determinism Introducing Factors sources and interactions in ML workflow}
    \label{fig:ndifs}
\end{figure*}

Fig. \ref{fig:ndifs} establishes a mapping between \ndif~and their interactions in ML workflow. When using a random number generator, the associated runtime will check to see if a special configuration was done which is called ``seeding'' the number generator. Providing a specific value for the seed should generate the same random number from the generator enhancing reproducibility. While a model is generally static in its definition, as an example from \dl, non-determinism of a neural network architecture can be showcased by the presence of some layers that are subject to randomness effects, like Dropout methods \cite{labach2019survey}. The main reason why a threading model can affect non-determinism is the non-associativity of floating-point operations. Therefore, training a model with multiple threads can affect the output of floating point calculations resulting in non-deterministic behaviour. Software versions encompasses two aspects: (1) direct software dependencies of the program and (2) software versions that are available in the runtime. Each component of a runtime environment may execute with different configurations affecting the output. As an example, a specific runtime might be built with support for double precision while another might only perform standard precision operations. The hardware used for a specific computation can yield different results in computational workloads, e.g., different instructions leading to inconsistent floating point precision. While using a specific dataset already split in training, validation and test sets, there is another factor to take into account: the inner ordering of each of the splits. For example, ordering of data affects the convergence rate of Stochastic Gradient Descent (SGD) as an optimization method in DL \cite{Meng2019}. 

On the other hand, ML frameworks (like PyTorch, TensorFlow and Keras) are widely used by ML practitioners in research and development and like other software these frameworks are prone to bugs. There are some empirical studies to understand bugs in ML but the focus of them is on bugs in client applications not bugs inside ML libraries. The only work that considers bugs in ML frameworks studied 202 bugs inside TensorFlow \cite{jia2020empirical}. While root causes and symptoms are analyzed and compared to bugs in traditional software, the empirical impact of bugs on the ML process/model was not touched \cite{jia2020empirical}. Quantifying the impact of defects in the process of \ml~allows us to make educated decisions on the versions used to develop and deploy a \ml-based software systems. Moreover, understanding the effect of defects in frameworks on the \ml~models will allow us to gain understanding on the robustness of models. As \ml~models are mathematical, if the defects affect operations within the model, the resulting overall equation might behave similarly while some of the ``inner'' operations might change. However, a prerequisite for such study is managing NDIF during ML training and inference. Otherwise, one can not be assured of the effect of a change (whether buggy or not) in the framework or model. In fact, an exact duplication of experiments is highly unlikely. The goal for \emph{reproducibility} is thus to reduce possible \ndif~in order to obtain the same results \emph{within the stated precision}\cite{methodologyDefinitionsACM}. Moreover, a proper calculation of the results with statistical analysis is paramount for reproducibility in \ml.

In this paper, we first introduce \framework{}, a toolset for minimizing the effects of NDIF in ML experiments (reproducible experiments). Then, we suggest a methodology to measure the impact of bugs in ML frameworks using \framework{}. Our contribution is two-fold: 1) introducing \framework, a framework for controlling some NDIFs, i.e., data split and random state configuration, in ML experiments, and 2) leveraging \framework{} to investigate the potential impact of bugs contained inside ML libraries, e.g. Pytorch, on training/inference of ML models. Although such NDIFs are well-known and should be managed in ML experimental designs, to the best of our knowledge, \framework~ is the first toolset that offers automated and systematic control of such factors during ML experiments. It facilitates designing experiments and provides the user with low-variance measurements by controlling NDIFs. The framework is completely open source and available online\footnote{\url{https://github.com/swatlab/ml-frameworks-evaluation}}.

The rest of this paper is organized as follows: Section \ref{sec:relatedwork} provides a brief summary of the related works. We introduce \framework~in Section \ref{sec:framework}, a framework for reproducible training and inference in ML. In Section \ref{sec:methodology}, we present application of \framework{} to investigate the impact of bugs in ML frameworks. To do so, we describe our designed methodology from data collection to statistical analysis. In Section \ref{sec:empirical}, we apply our methodology on \pytorch~and present the results including statistical analysis. In Section \ref{sec:limitations}, we explain the main limitations and threats to validity of this study. We conclude the paper in Section \ref{sec:future-works} by explaining potential future research directions.



\section{Related works}\label{sec:relatedwork}
Vanschoren \etal~tried to standardize and centralize research results and processes for \ml~by constructing ``experiment databases''\cite{vanschoren2012experiment}, allowing researchers to explore and query results from \ml~findings. Building on this work, OpenML is created, a platform allowing for sharing datasets, implementations and results among \ml~researchers \cite{vanschoren2014openml}. Using such platform has the benefit of improving reusability and reproducibility of previous \ml~research. In another work, authors set the same random seed in order to measure difference in performance between different random seeds, and between \pytorch~and TensorFlow \cite{jean2019issues}. They show a 7\% difference between the random seeds; which is similar to the aforementioned results. More interestingly, they look at the ``Sensitivity to Data Ordering'' and show that there is as much as 3\% difference when changing the order in which the data was fed.

There is a promising tool for managing the \ml~workflow in an ecosystem called MLFlow \cite{zaharia2018accelerating}. Its main goal is to simplify the lifecycle of a \ml~project by wrapping over existing processes. Such encapsulation of the lifecycle, they argue, benefits the \ml~practitioner. Moreover, reproducibility is boasted as one of its four challenges addressed. MLFlow allows for tracking of experiment metrics and configuration via file configurations.

A survey was conducted by Isdahl comparing the existing \ml~``platforms''. These platforms are ecosystems for managing the \ml~workflow in some way or another \cite{Isdahl2019}. They rank these platforms by comparing their ``out-of-the-box reproducibility''. They show that most of these ecosystems offer some degree of reproducibility, but none of them achieves total reproducibility. Therefore, reproducibility is not yet completely managed by a single \ml~ecosystem and it is up to practitioners and researchers to make sure their work is reproducible.

Data Version Control (DVC) is a tool that aims to facilitate scientific experimentation in \ml, both in terms of shareability and reproducibility \cite{Kuprieiev2020DVC}. They use a variety of techniques and tools to make sure experiments can be easily shared and reproduced. They version data, models by using configuration files, without tying themselves with a specific run configuration, thus making it framework-agnostic. Our work, \framework, resembles DVC closely as we also use configuration files to control reproducibility, notably for datasets. Our approach differs in that it acts more as a central database for collection and reproduction of runs. Actually, \framework{} may benefit from DVC in order to verify that data stays the same between experiments. Moreover, we aim at ensuring that the \emph{runtime} on which experiments are executed are tightly controlled, in order to collect measurements. Our work can benefit from DVC in order to make sure data stays the same between experiments.

Other tools such as COBRA \cite{Vogtlin2020cobra} aims to create reproducible setups of workflows by installing tools via a command line interface. A tool called ReproZIP is meant to replicate experiments across various computers by capturing dependencies using tracing technologies \cite{Chirigati2016reproZip}. They argue that configuration and moving from a research environment to a computational environment requires manual changes that discourage users from creating reproducible setups. There are certain ecosystems that have been developed for reproducible computations in the cloud, namely MULTI-X \cite{Vila2018}, PRECIP \cite{Azarnoosh2013}, and other unnamed solutions \cite{Orzechowski2020}. 
\section{The framework}\label{sec:framework}
The creation of \framework{} is motivated by the need for reproducibility in \ml~and the lack of existing solutions at the time of creation of the framework. It was originally designed to reduce the amount of variance between multiple runs of ML experiments (training or inference).

At its core, \framework{} is a training benchmarking framework developed by \Python{} focused on \emph{repeatability} and striving for \emph{reproducibility}. As there are technical limitations, prohibiting complete reproducibility, we will use the term ``reproducibility'' to indicate both ``repeatability'' and ''reproducibility''. It is important to note that \framework{} is not a \ml~framework: rather, it configures the necessary runtime mechanism for other existing \ml~frameworks in order to have reproducible results. It focuses on reproducible data splits and reproducible random state configuration where possible. Another goal of \framework{} is to have a simple usage with as simple as possible opt-in mechanism: this is done by defining clear interfaces. While \framework{} is young and currently only supports \Python{} and \pytorch{} for its runtime configuration, additional configuration for other \ml~frameworks is well within the realm of possibility. It is primordial to consider that \framework{}~aims to address reproducibility at the data and randomness aspects of the ML workflow. The goal here is not to create an ecosystem to ensure reproducibility but rather to provide a framework for the specified parts: runtime configuration, data management and randomness seed archival. Other tools exist and their combined usage with this framework is what will allow empirical studies to achieve better reduction of confounding factors. \framework{} performs the concept of ``data versioning'' but has some key differences to DVC \cite{Kuprieiev2020DVC}: our work is a benchmarking tool that integrates data versioning. DVC, however, is to be used in the \ml~workflow.

\framework{}, as a benchmarking system provides a centralized and clearly defined metric collection procedure. In short, it consists of two portions: a client and a server, each of which needs special configuration to achieve reproducibility. The client trains \ml~models in tightly controlled experiments which are given by a server. The server is responsible for giving the training and test data subsets in a deterministic fashion and to collect results from multiple clients. The server can be viewed as static, whereas the client is dynamic. Indeed, the server acts as a central repository for data collection, challenges and most importantly deterministic data serving to clients. Clients, on the other hand, are expected to widely vary in configuration, hence there are two key aspects to consider: (1) reproducible random seed initialization and processes during training and (2) communication with the server in order to collect data and send metrics. What makes this portion of the framework stand out from most other available frameworks is its focus on reproducibility rather than performance. We chose a client-server model where \framework{}’s server is implemented as a metric collection center to allow reusability and clients are expected to widely vary in runtime configuration. In this section, we explain how \framework{} deals with the essential parts of the ML workflow in order to have better reproducibility and repeatability than standard \ml~workflows. 
\subsection{Client}
The client interfaces with libraries and has the responsibility to configure the runtime using reproducibility mechanisms available in the dependencies. Therefore, the client is strongly tied to ML frameworks and their versions. The client uses the built-in mechanisms of dependencies and frameworks to correctly set the runtime. As an example, the core of the framework comes with support for \pytorch{} and interfaces with the runtime for model training. Interoperability with various frameworks is done by leveraging the interpreted nature of \Python{} and its mechanism for dynamic imports. Using this, \framework{} sets the several mechanisms to make sure the randomness is controlled.

Users are expected to import their models by inheriting from a base class or creating a class that contains the model. The training procedure is entirely controlled by the framework as the interface from the base class is representative of a typical \dl~training procedure. Currently, \framework{} comes with two predefined models for image classification: VGG \cite{simonyan2014very}, and AlexNet \cite{krizhevsky2017imagenet} taken from the official implementation of \pytorch{}'s related module Torchvision \cite{torchvision}. It also comes with a \CNN~inspired from VGG but adapted for working with 32x32 images, internally named \texttt{VGG-X}. When running a model with the framework, one specifies the runtime she wants to load and the model name, among other parameters. Therefore the goal is for a user to write her model with the \ml~frameworks of her choice, make sure it is supported by the framework and launch a client running procedure. Provided that the server is running, this setup will allow for reproducible runtime configuration done by the framework and the training sequence can begin.
\subsection{Server}
The server responsibility of \framework{} is two-fold: (1) keeping metadata and metrics about experiments and (2) serving data to a client in a deterministic and reproducible fashion. The server controls the metadata by holding and safekeeping seed values for an experiment. In this context, an experiment is a unique identifier that is supplied by the client. Deterministic random seeds are generated when a new identifier is transmitted by a client and are saved in a file by the server. These seeds are used both by the client and the server to configure the randomness on their end. On the server-side, reproducible data splitting directly uses the random seed allowing for consistent serving of data. Moreover, the server collect metrics from runs by establishing a session with a simple protocol for communication. This protocol is followed by the current \Python{} implementation provided in \framework{}.

It is also possible to use the server standalone and simply leverage the communication protocol defined within \framework{}, as the underlying protocol is TCP. Therefore \framework{} could be used to make a simple metric collection server and dashboard for comparison based solely on the identifier. This allows for others to provide similar work of controlling randomness on the clients not using a \Python{}~runtime.

\subsection{Communication}
The client and the server communicate over TCP to exchange training data, seed information and metric collection.
There are some rudimentary safeguards currently implemented in \framework{}'s communication protocol to make sure data is correctly received and unchanged. Notably, as a precautionary action, both parts of the communication check data integrity when receiving information.
Security of the communication channel was not studied nor implemented in the context of this work as the current communication is meant to be done within trusted machines and/or networks.

Another precautionary action taken is by sending the value of the seed when the client asks for data for a specific run. The seed is compared to the server's version and if a mismatch is found, the server stops. Moreover, a checksum is used on both sides to make sure data integrity and ordering are preserved.

\section{Impact of bugs in ML frameworks: A case study}\label{sec:methodology}

Our research question is this section is as follows: \textit{Given a sample of fixed size of recent bugs in a given ML framework, does the occurrence of bugs affect the performance metrics of ML models?} In the following, we describe a general methodology applied in order to measure the impact of \emph{code changes} that intended to be bug fixes in a \ml~framework using \framework{}. The main idea is to obtain versions of a framework before and after a change is introduced and measure the impact of the change based on a particular evaluation procedure. 

We define a \emph{change} as the difference between two accessible \emph{versions} of a software artifact. Practically, a change is a difference between \emph{versions} in a Version Control System (\VCS), \ie a difference between two ``commits'' in \git. The presence and absence of a bug, by the process of bug fixing can be seen as a change: therefore this methodology applies directly to bugs; since the overall idea is to quantify the impact that the presence or absence of a bug has on ML model performance, we use two different versions of the framework. The first version containing the bug is denoted as ``\buggy{}'' whereas the version that does not have the bug is denoted as ``\corrected{}''. Generally, for an end-user of a library, the \buggy~and \corrected~versions would be separated by an official release of the software. However, while applying the methodology using official releases would be much faster and accessible, by doing so we would not measure \emph{only} the bug presence, but also all the other changes that were bundled in the same version. Therefore, we opt not to use official releases. We rather compare the absence and presence of a bug by obtaining an artifact, \ie~a build of version, with and without the defect's presence. 
Consequently, for each bug the key information is its \buggy~and~\corrected~versions. In the current study, we propose to set the ``\buggy''~and~``\corrected''~versions as respectively, the \emph{last version} that contained the bug and the version that contains the bug fix. Therefore, the bugs are separated by a single commit \label{methodology:keypoint:change}.

The general process consists of the following steps: (1) Data collection, (2) Data Filtering, (3) Artifact generation, (4) Evaluation (experimental runs) and (5) Statistical analysis. These steps are generic and can be applied to any \ml~framework, however special considerations might lead to small experimental differences from framework to framework. The methodology is designed to reduce the amount of confounding variables and non-determinism, and thereby variance. In the rest of this section, we will present the details of each step. 



\subsection{Data collection}\label{methodology:data:collection}
The first step is to collect information about the past defects in the framework, \ie~identify what bugs are present. There are various ways this can be done as a consequence of ML frameworks varying in size, software engineering complexity, documentation and release process.
In either case of manual or systematic bug collection, the core information that needs to be available at the end of this step is: (1) \mversions~versions of each bug, (2) enough information to build an artifact for each version and (3) enough information on the understanding of the bug to apply a decision on the filtering process. As mentioned before, we take \mversions~versions to be before and after a bug fix is applied to the main branch of the project. In the following, we present specific methodological considerations taken in order to carry out an empirical study for the bugs in \pytorch{}.

The study analyzes the bugs on specific releases ranging from version \version{1.1.0}~up to version \version{1.6.0}. 
In the case of \pytorch{}, the software release process has been streamlined and changes are usually correctly reported in their \textbf{Release Notes} with corresponding  bug fix information. Therefore, we use the bug fix sections of each release note to construct our corpus. We followed this methodology since the bugs that are reported by the maintainers are guaranteed to be coming from the framework and are definitely labelled as bugs. As \pytorch{}'s repository is hosted on GitHub, it is possible to mine the issues to find the bugs since GitHub provides a web \API{} to get information about issues opened and their labelling scheme. However, we found that in practice this approach leads to more complications, as a bug might be wrongly labelled by the triage system in place. Additionally, using an issue-mining approach is time dependent, as labels can be added and removed at any time. Following the guideline of never changing the content of an official release by using consistent semantic versioning principles allows us to mine the information directly from the release notes. We extracted \nBugsCollected~bugs following an automated approach to parsing of release notes published by the maintainers. This parsing yielded metadata such as the \PullR{} number, links to description, etc. We call this set of all collected bugs as $B_a$.
\subsection{Data Filtering}\label{methodology:data:filtering}
After having collected all the bug information, there is a necessary step of manual filtering, \ie creating a subset of $B_a$ which is called $B_f$ containing bugs that are relevant (independent of end-user) and ``silent''. Each project being managed differently, specific methods, adapted for each framework's development life-cycle, to list potential bugs to be studied are to be considered. However, for each framework, the following conditions for a specific bug to be examined share the same rejection criteria. For each bug $x \in B_a$:
\begin{enumerate}\label{methodology:steps:filtering}
    \item Reject $x$ if it causes a compilation error,
    \item Reject $x$ if it causes a runtime error, causing the application to exit (crash),
    \item Reject $x$ if it is caused by end-user code.
\end{enumerate}

Our reasoning behind these criteria is to only study bugs that are \textit{silent} and could easily be missed during the process of training \ml~models. In the first two cases, we therefore remove all non-silent bugs. Lastly, we do not consider bugs that would be opened due to errors by end users, as they do not constitute a bug of the \ml~framework itself. Bugs that comply with the such conditions are added to the set of issues to be examined, $B_f$. We left with \nBugsLabelled~bugs at the end of this phase.

For the current study on \pytorch{}, we added the following constraints to limit the search of bugs. These constraints were applied in order to prioritize bugs that would have a clear impact on training and that would be easy to replicate.
Thus, a manual filtering effort was made with the following additional criteria:
\begin{itemize}
\item Rejection criteria:
\begin{enumerate}
    \item The bug is on CPU, so running an experiment may take too much time in comparison to GPU,
    \item We cannot find an application that would be affected: the bug must have an impact on ML experiments: creating a ML model or using it for training/inference.
\end{enumerate}
\item Favourable acceptance:
    \begin{enumerate}
        \item Related to gradients,
        \item Related to mathematical functions,
    \end{enumerate}
\end{itemize}
We use ``favourable acceptance'' instead of ``acceptance'': this is done to avoid false negatives. This effort was made by three persons possessing a working knowledge of \ml~and \dl. This criterion still may lead to false positives and false negatives that is discussed in Section \ref{sec:emprical:discussion}. We analyzed \nBugsLabelled~bugs out of which \nBugsUsedForStudy~met our criteria.
\subsection{Artifacts generation}\label{methodology:artifacts}
This step aims to create artifacts that can be used to measure the impact of the bug at both \mversions. For each bug-revision pair, one needs to go through the entire build pipeline to produce an artifact that is suitable to be installed and evaluated. The build process between a bug's \mversions~revisions have to be identical, while the build and linking dependencies need to be tightly controlled. Ideally, build dependencies are identical as well as the build process. While this specific step is to-the-point, there are magnitude of precautions and limitations that one needs to take into account. Section \ref{sec:limitations} discusses the precautions and limitations in more detail, notably on why a single build configuration environment is probably not enough. Depending on the framework under study, certain other precautions might apply. However, for each bug $x \in B_f$, we:
\begin{itemize}\label{methodology:steps:artifacts}
    \item Identify the build tool chain and configuration needed, namely $T_x$,
    \item Build a version $x_b$ for the \mbuggy~version using $T_x$ at revision \mbuggy,
    \item Build a version $x_c$ for the \mcorrected~version using $T_x$ at revision \mversions.
\end{itemize}

We successfully completed the build process for the buggy and corrected versions of 49 bugs out of 115. For the rest, we could not successfully run the entire build pipeline for all necessary versions due to issues in building and linking dependencies.

\subsection{Evaluation (experimental runs)}
This step aims to define the measurements of performance of ML models running on artifacts, for both \mversions~versions.
The general idea is to train both versions on the same model, hyperparameters, data, randomness configuration, data dependencies and hardware and gather a series of measurements during training and evaluation: these are \emph{experiment attributes}.

An \emph{experiment} constitutes the whole metric collection process of a training procedure executed multiple times, each repetition being denoted as a \emph{run}, for a specific version of the framework (Table \ref{methodology:exp:epoch-run-exp}). In this context, a training procedure aims to recreate as closely as possible the process of learning used in modern ML; a run's output is a trained model that gives acceptable performance along with the metrics calculated against a test set. An \emph{experiment} is uniquely characterized by the attribute in Table \ref{methodology:exp:char}. More specifically, a training procedure (run) is the set of steps taken to train a model from scratch. In the case of neural nets, the training procedure consists of multiple training epochs over the same dataset and updating the weights. For each bug, we will conduct two experiments: one for the \mbuggy~and one for the \mcorrected. We use the term \emph{performance} as various measurements of a model's capability to correctly accomplish its task. Notably, for classification problems, metrics used are Accuracy, Precision, Recall and F1-score, among others.

For a single bug $x \in B_f$, let $e_b$ and $e_c$ be its \emph{experiment} for respectively \mbuggy~(b) and \mcorrected~(c) versions. The only experimental attributes (see Table \ref{methodology:exp:char}) that differ between $e_b$ and $e_c$ are their \emph{evaluation type} $t$ and artifact $b$.
The following procedure is done to collect metrics for an experiment consisting of $N$ runs, it characterizes a model's performance given a specific set of experiment attributes:
\begin{itemize}\label{methodology:steps:analysis}
    \item At the start of each run $r_i$, all random configurations must be reset to specific state $s$,
    \item Each $r_i$ must consist of a training algorithm spreading over $e_t$ epochs,
    \item At the end of each $r_i$, measure performance metrics on test set and record metrics,
\end{itemize}

An appropriate number of runs $N$ is to be set for analysis: this depends on various factors, but we recommend at least 30. Hence, for each bug $x \in B_f$
\begin{itemize}\label{methodology:eval:steps}
    \item Choose experiment attributes $e$ (Table \ref{methodology:exp:char}) representative instance of $x$
    \item Apply evaluation procedure with experiment attributes for $e_b$, the \mbuggy~version.
    \item Apply evaluation procedure with experiment attributes for $e_c$, the \mcorrected~version.
\end{itemize}

Table \ref{methodology:metrics:sample} shows a sample output for a single experiment. We attempted to run experiments with all 49 bugs and manually inspected the results. The goal was making sure that experiments were out of error (e.g., failure of training procedure) and conducted under the same experimental setup. This process led to the exclusion of 39 bugs and finally, we were able to complete our statistical analysis on 18 bugs.

\begin{table}[t]
\centering
\caption{Composition of experiment steps}\label{methodology:exp:epoch-run-exp}
\begin{tabular}{ll}
    Epoch   & Ran multiple steps \\
    \hline
    Run     & Ran multiple epochs\\
    \hline
    Experiment  & Ran multiple runs\\
    \end{tabular}
\end{table}

\begin{table}[t]
\centering
\caption{Attributes of an experiment}\label{methodology:exp:char}
\resizebox{1\columnwidth}{!}{
\begin{tabular}{lrr}
    Bug identifier & - & A user given name \\
    Evaluation type & $t$ & Either ``\mbuggy'' or ``\mcorrected'' \\
    Model & $m$ & \underline{M}odel used to train \\
    Challenge & $c$ & The dataset used to train the model \\
    State & $s$ & The random \underline{s}tate used \\
    Artifact & $b$ & Specific \underline{b}uild that showcases the bug \\
    Software & $d$ & Software \underline{d}ependencies used for the runtime \\
    Epochs & $t$ & Number of epochs used to train the model \\
    \end{tabular}
    }
\end{table}

\begin{table}[t]
\centering
\caption{Metrics collection sample}\label{methodology:metrics:sample}
\begin{tabular}{lrrrr}
 & 0 & 1 & ... & $N$\\
\hline
\textbf{accuracy } &  $a_0$ &  $a_1$ &  ... &  $a_n$ \\
\textbf{precision} &  $p_0$ &  $p_1$ &  ... &  $p_n$  \\
\textbf{recall   } &  $r_0$ &  $r_1$ &  ... &  $r_n$  \\
\textbf{f1       } &  $f_0$ &  $f_1$ &  ... &  $f_n$  \\
\end{tabular}
\end{table}

\subsection{Statistical analysis}
In order to quantify the impact of each bug, two distributions are measured: the distributions of performance metrics when the bug is present (\buggy{}) and when the bug is corrected (\corrected). To reduce the variability, the same random seed is used for each experiment, that is both the \mversions~versions of a bug will be trained using exactly the same data and initial random seed. More generally, the \mversions~versions should run with identical \ndif{}.

In order to answer our research question, we define the following hypotheses:
\begin{itemize}
    \item[$H_0$] For a training procedure, there are no differences in a metric between the \buggy{} and the \corrected{} version of a bug fix.
    \item[$H_1$] For a training procedure, there is a difference in a metric between the \buggy{} and the \corrected{} version of a bug fix.
\end{itemize}
We formulate the alternative hypothesis as a strict inequality since we have no prior knowledge of measurements, \ie~we use a two-tailed test. We use \wmw~test between the performance metrics of each version, \ie between the training procedure results using $e_b$ and $e_c$. \wmw~is a nonparametric test that measures the difference in mean ranks between two samples \cite{Bergmann2000}. There are two versions of this test: one in which observations are paired and one that observations are not paired. Since all initial factors are identical for each run, ordering of the runs is not important. Furthermore, the samples are not considered paired, as each run contains the same random seed configuration and is reinitialized to its initial state. In this case, as the metrics for runs are not paired we use the non-paired \wmw's test: the \emph{Wilcoxon-Mann-Whitney U-test} with a 95\% confidence interval for measuring the impacts.

Regarding the experimental setup, one needs to train models repeatedly to get all the required performance metrics. However, using a traditional environment for training does not suffice. Indeed there are a plethora of factors affecting the performance of a \ml~model. In this study, it is imperative that the runtime and its dependencies be the same for both experiments of the same bug. Therefore, a consistent runtime needs to be used: we employ \framework{}, presented in Section \ref{sec:framework} in order to consistently collect the data needed for statistical analysis. While \framework{} deals with the configuration of randomness and data across runs, there are other mechanisms that it cannot control, notably the runtime itself. Consequently, a virtual machine with predefined Docker \cite{Docker} images provides a good compromise in runtime configuration and ease of configuration.

\section{Execution and results}\label{sec:empirical}
In this section, we present the results of application of our methodology on a set of bugs in \pytorch{} using \framework{}.

\subsection{Experimental setup}\label{sec:empirical:execution:setup}
In the following, we overview the overall experimental configurations. For a complete list of our hardware and software used throughout the sections, please see the online repository\footnote{https://github.com/swatlab/ml-frameworks-evaluation}.

To build the various versions of \pytorch{}, we used Google Cloud Platform (GCP) virtual machines running \Docker~containers. We use two images to build the versions: the first is a container loaded with the tools needed to build \pytorch{} at \Python{} \version{3.7.9} while the other is for the version \version{3.6.7}. This dual setup is to make sure that we replicate a plausible build. For running the experiments, we leverage virtual machines for hardware control and Docker containers for software and runtime dependency control. The GCP virtual machine used is a \texttt{n1-standard-16} with virtualization on \emph{Skylake} \CPU{}s running Linux kernel version \version{5.4.0-1028-gcp}, on Ubuntu \version{18.04.5} equipped with a \emph{NVIDIA V100-SXM2-16GB} running driver \version{450.51.06} with \CUDA~\version{11.0}. Experiments ran on Docker \version{19.03.13} accelerated by the nvidia-container-runtime.

Our current experimental runs use \texttt{VGG-X} from the default models provided by \framework{} to run experiments on working on the CIFAR-10 dataset \cite{CIFAR10}. This choice is motivated by a good balance between complexity, time to train, and relative novelty of the model. We run our model for 30 epochs over 50 experimental runs for each of the \mversions~versions. The optimizer is set by the framework and uses SGD with a learning rate of 0.01 and a momentum of 0.5. We chose to experiment on a Docker container in order to provide a reliable and reproducible software configuration. The overhead induced by Docker is not noticeable and has been shown to be small when training models \cite{Xu2017Docker}. Table \ref{experiment:attr} shows the attributes we use for our experiments. The random state used varied from a bug to another, as we are not interested in measuring the stability for a single random seed. Within a single bug, the random seed stays identical for all the runs.
    
    \begin{table}{}
    \centering
    \caption{Shared attributes for our runs}\label{experiment:attr}
    \begin{tabular}{|c|c|}
        \hline
        Model & VGG-X (no Batch Normalization) \\
        \hline
        Challenge & CIFAR-10 \\
        \hline
        State &  \textit{Depends on the bug} \\
        \hline
        \multirow{2}{*}{Software} & \textit{Depends on the bug,}\\
        & \textit{but Docker image contains specifics}\\
        \hline
        Epochs & 30 \\
        \hline
    \end{tabular}
    \end{table}
    
Since our challenge is CIFAR-10, the problem is multiclass. Therefore, there are multiple ways to calculate the accuracy, precision and recall. In this study, the scores reported are using the unweighted mean of scores of each metric. That is, we average the metric for each class, without weighing by the number of representatives in each class.

\subsection{Results}\label{sec:empirical:analysis}
We have been able to measure successfully \nBugsResults~bugs and conduct the \wmw~test on their \buggy~and \corrected~versions. 

We present the results after having processed the log files emitted by our experimental setup. There were manual steps needed to parse and verify that results were clean of any mistakes. Notably, we verified that each experiment was conducted under the same experimental setup as described in Section \ref{sec:methodology}. There are two bugs that did not complete the 50 runs mandated for both \mversions, namely experiment \texttt{study-pr36832} for its \buggy~version completed 47 runs (and 50 for its \corrected). The second is experiment \texttt{study-pr31433} which completed 43 runs for its \corrected~version and 50 for its \buggy~version. All other bugs completed 50 runs; we mark with ``\dag'' results that did not complete full 50 runs for both versions. We kept these two bugs for full disclosure.

As mentioned in the methodology, we use a \wmw~non-paired test to indicate whether or not the results did show an impact. Table \ref{table:empirical:pvalues} shows the p-values of the hypothesis test for all of our experiments. The \wmw~test can accept a different number of samples in each population, but in this case, we choose to only compare the same number of samples in the population. That is, for the bugs that did not complete the full experiments, the values portrayed are for a \wmw~test done with the minimal number of runs available by either \buggy~or \corrected~version.

\begin{figure}[t]
    \centering
    \includegraphics[width=\linewidth]{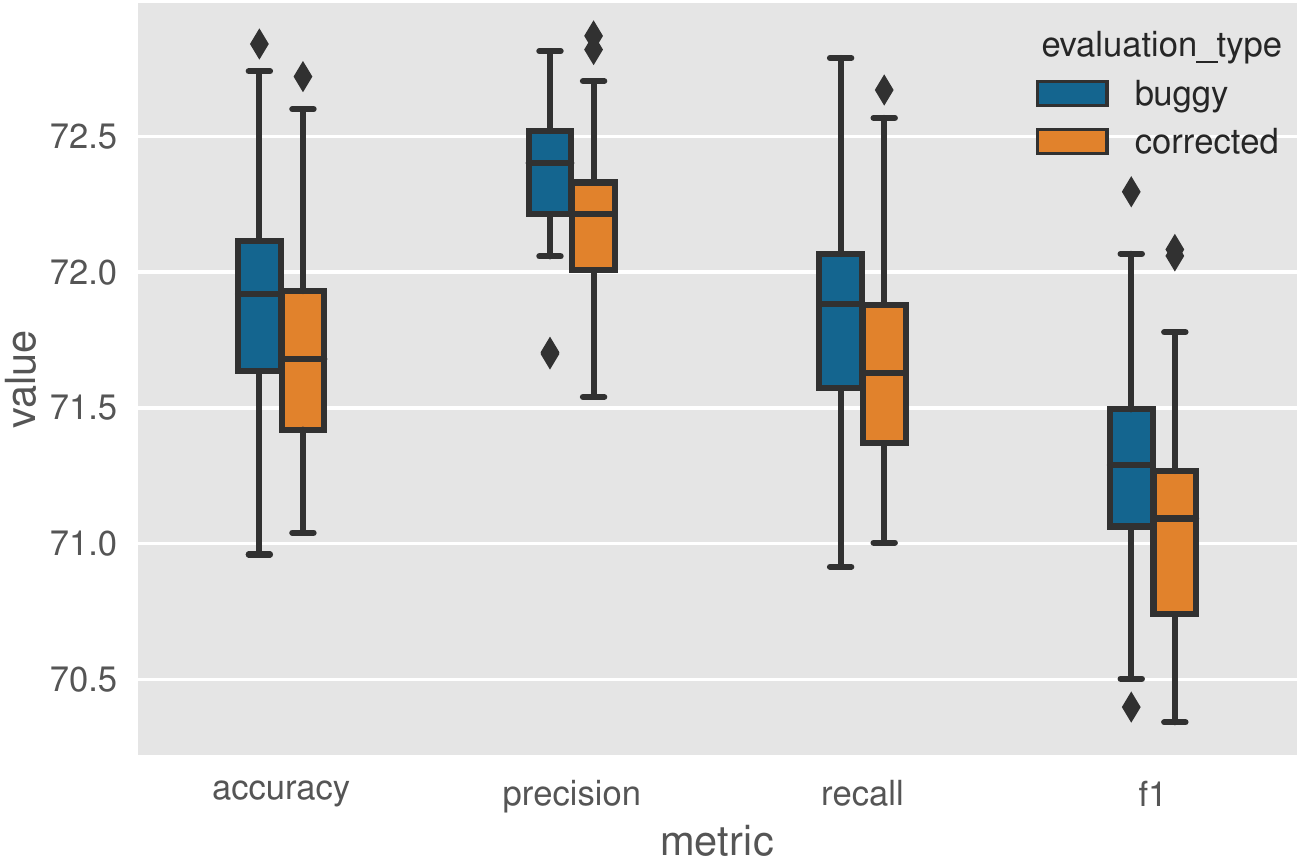}
    \caption{Distribution of metric values for bug \bugname{study-pr31433}}
    \label{fig:empirical:31433}
\end{figure}

We can see that only one bug did show in fact a statistically significant result for our experiment: \texttt{study-pr31433}. This bug's \corrected~version only completed 43 runs. Even when computing the p-value by using the maximum number of samples in each population (50 for \buggy~and 43 for \corrected), we still obtain p-values less than 0.05 for each metric. We can see the distribution of each metric in Fig. \ref{fig:empirical:31433}. Interestingly, this bug is not related to our \CNN: the bug fix is related to completely different files, \ie LSTMs. We categorize this value as an outlier as there is no reason to suspect there was a difference between the two versions. 
We can therefore say that with the current data, there is no evidence to reject $H_0$, thus we cannot conclude that studied bugs have an impact on model performance. We will discuss several aspects that introduced non-determinism in our experiments in Section \ref{sec:limitations}.

\begin{figure}[t]
    \centering
    \includegraphics[width=\linewidth]{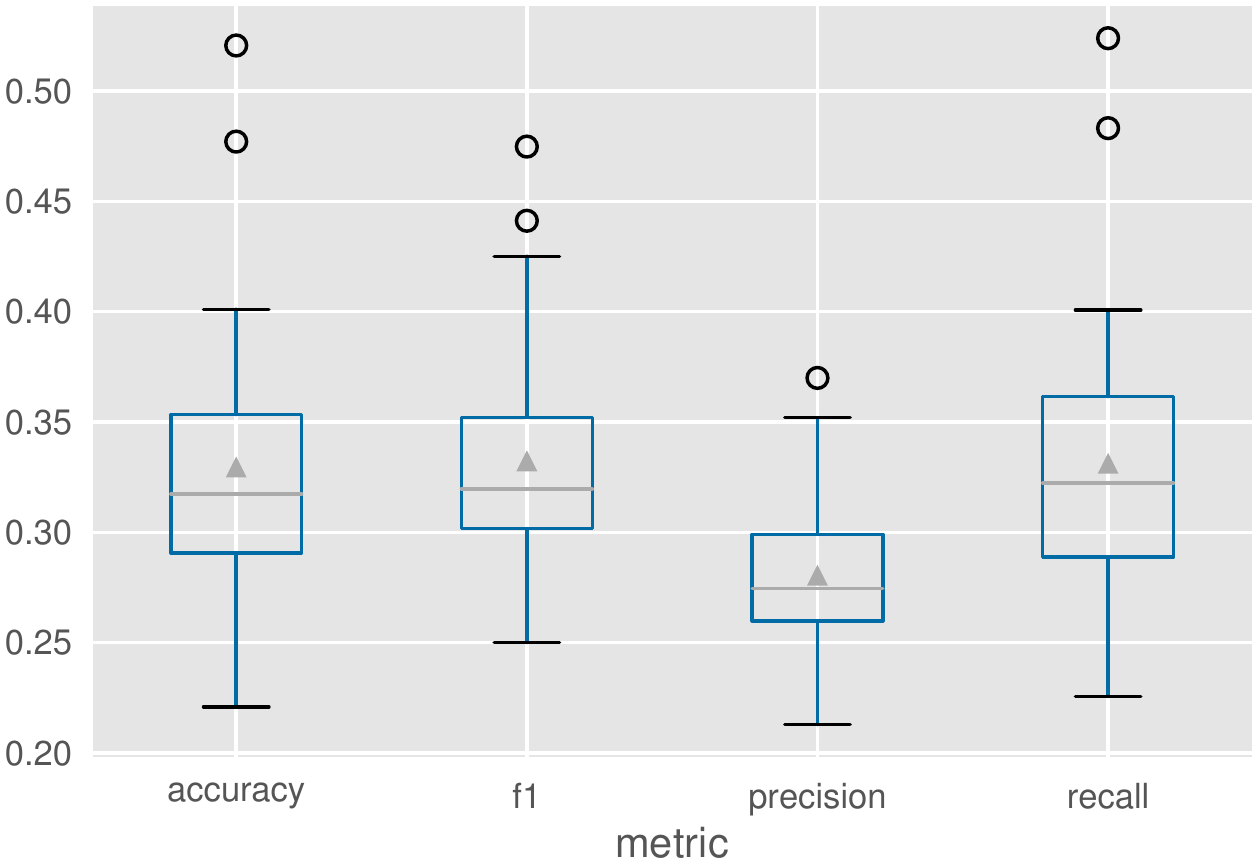}
    \caption{Distribution of metric standard deviations across all experiments}
    \label{fig:empirical:stddev}
\end{figure}

We here report the standard deviation of our runs per metric. This allows us to observe the effects of \ndif~on the runs. We can observe in Fig. \ref{fig:empirical:stddev} that the standard deviation for each metric is pretty small for all runs, indicating that some control of \ndif~was done, without, however, quantifying the impact that is left to control. 

\begin{table}\caption{p-values of the hypothesis test for all experiments}\label{table:empirical:pvalues}
\centering
\resizebox{.9\linewidth}{!}{
\begin{tabular}{@{}lrrrr@{}} 
\phantom{abc} & \multicolumn{4}{c}{Metrics}\\
\cmidrule{2-5}
&     Accuracy &           Precision &    Recall &       F1 \\
\midrule
study-pr31167    &  0.62929 &   0.42190 & 0.58365 & 0.88217 \\
\textbf{study-pr31433}\dag    &  \textbf{0.03320} &   \textbf{0.00243} & \textbf{0.03508} & \textbf{0.01057} \\
study-pr31552    &  0.92581 &   0.58365 & 0.97525 & 0.73292 \\
study-pr31584    &  0.58356 &   0.56486 & 0.62208 & 0.57893 \\
study-pr32044    &  0.84956 &   0.39074 & 0.78009 & 0.48842 \\
study-pr32062    &  0.46259 &   0.87673 & 0.45033 & 0.77481 \\
study-pr32350    &  0.48821 &   0.55096 & 0.53270 & 0.65657 \\
study-pr32541    &  0.63418 &   0.63185 & 0.61237 & 0.39074 \\
study-pr32829    &  0.50135 &   0.21085 & 0.50589 & 0.39455 \\
study-pr32831    &  0.66144 &   0.94230 & 0.65657 & 0.95327 \\
study-pr32978    &  0.31404 &   0.29949 & 0.35740 & 0.44620 \\
study-pr33017    &  0.14363 &   0.21849 & 0.12005 & 0.10014 \\
study-pr35022    &  0.28821 &   0.51031 & 0.29629 & 0.32932 \\
study-pr36820    &  0.20198 &   0.53270 & 0.20097 & 0.57422 \\
study-pr36832\dag    &  0.30004 &   0.37223 & 0.26631 & 0.26958 \\
study-pr37214    &  0.67396 &   0.72257 & 0.69181 & 0.69181 \\
study-pr38945    &  0.40400 &   0.97525 & 0.36465 & 0.58365 \\
study-pr39903    &  0.97249 &   0.74854 & 0.98625 & 0.93681 \\
\end{tabular}
}
\end{table}



\subsection{Discussion}\label{sec:emprical:discussion}
For each experiment, the standard deviation for all metrics across runs is situated between 0.2\% and 0.51\%, showing that controlling \ndif~was done relatively well. We compare ourselves to work done by Pham \etal \cite{Pham}, in which we can observe that the larger models have a bigger standard deviation than smaller models when using a fixed seed. Notably, our model uses around 15 million parameters with the Dropout effect encountered and still shows mean standard deviations around 0.3\%. Comparatively, for Pham \etal's most comparable model in number of parameters, \ie~ResetNet56 shows similar range, around 0.5\% of standard deviation. While there is no hard evidence of the effect of size, \ie statistical test to show this, based on observations we think that bigger models have a higher potential of being affected by \ndif. It would be interesting to see how Pham \etal~constructed their models, since they said it was reviewed multiple times \cite{Pham}. However, at this time, their repository does not seem to contain their implementation\footnote{\url{https://github.com/lin-tan/dl-variance}}.

During our experiments, there were \nBugsExcluded~bugs that have been excluded for not using the same seed: these are the consequences of ungraceful shutdowns due to GCP specificities, leading to data loss of seed value. While these results were discarded for not using the same random seed, they did prove statistically significant impact. Consequently, we can argue that when using different random seeds, a statistical test such as \wmw~might positively reject $H_0$. 




\section{Limitations and threats to validity}\label{sec:limitations}
This section exposes the different limitations that are present in our realization of the methodology from a theoretical and practical standpoint.

\subsection{\framework~framework}
To add a new \ml~framework, the main hurdles to usability is that \framework{} is limited by the external \ml~frameworks' ability to control randomness and ensure reproducibility. Therefore, we strongly encourage \ml~framework developers and hardware makers to provide compatibility for reproducible workflows. One limitation of the current implementation of this framework is the backup mechanism and checkpoint system for seeds: there is a mechanism in place in order to save the values when the program ends or a signal is received, but this needs to be done more periodically, perhaps at each generation, as unexpected errors can cause seed values to be lost in environments where programs can end at any moment. Moreover, as of now there is no validation set generation, \ie~data splits are made in ``train'' and ``test''. As the primary intended use was for controlling specifically the random seed for the methodology, and since validation sets are used to tune hyperparameters and provide cross-validation, this feature was not included. It is however trivial to add support for validation sets if one desires to use it as a metric collection and challenge database. There are multiple factors affecting the stochastic aspect of training a model that cannot be controlled by runtime configuration: \eg~installed version dependencies and the runtime itself. These issues are meant to be addressed in future work or by other tools and techniques.

\subsection{Other NDIF}
There are multiple non-deterministic sources that have not been controlled. Our work is currently limited and does not offer any guarantee against multiple \cfvars~such as floating point calculations, possible thread interactions and \gpu~models. These are however mitigated by the fact that we carry the computations by using the same hardware, runtime, software versions and threading model for our experiments. Consequently, the results for a specific bug, \ie~between a \buggy~and a \corrected~experiment, use the same experimental attributes, thus reducing the effects of \ndif.

Specifically for thread interactions, \framework{} uses \pytorch{}'s built-in mechanisms in order to control its thread interactions. So, our experiments use only one \CPU~core to do its work, thereby eliminating some thread interactions from sources of errors. However, there is not a guarantee that \pytorch's thread interactions are complete, and our work is based on their suggestions for reproducibility \cite{PytorchRepro160, PytorchRepro110}. By detailing the current running hardware and related configurations, our work should be repeatable.

While we use \framework~in order to control the random seed on runtime, there are still variations present in the runs. There are multiple sources for this randomness that have not been controlled throughout the experiment. This indicates that additional precautions need to be taken when interpreting results. We believe that they are the result of \ndif~at the implementation level, as results by Pham \etal show a similar range of values for a similar model to ours \cite{Pham}.

Our current work does not currently support detecting if floating point operations are the same between experimental runs. There has been some research on making float point operations more stable \cite{Koester2017flexpoint}, but we are aware that our current framework values only hold for a specific architecture, which is why we disclose the hardware and software used in the study. Furthermore, adding support for such tools would most likely introduce more challenges and sources of error, as they are not integrated by default within \ml~frameworks.



\subsection{Triggering the bug}
The first and most important concern in this study is validating if bugs were actually triggered during the execution of our runs. We are aware that this is a huge limitation, as the core of this study relies on it. While the bugs could be triggered during manual execution, \eg by reproducing the bug with the code from the pull requests, this does not mean that when processing data in a training process this would happen. This is due to multiple factors, but the most important is that \pytorch~is a big framework with multiple ``backends''; here, the key is building the right version, with the right build parameters to make sure that the bug is triggered. Indeed, if the bug was not triggered, this study only means to demonstrate the variability induced by \ndif~even by controlling the seed, for within bug experiments. It can also be used to show the variance between multiple random seeds. It should be noted that on a very small scale, since most of the runs (inter-bugs) did run on different seeds, the split and the ordering of the data fed might affect the final training efficiency.

\subsection{Model and dataset used}
One might wonder why only a variation of VGG was used throughout this experiment. Indeed, there is no way of knowing during training if the workload contains the bug, as it is silent. Therefore, instead of using multiple models to accomplish the same task, we use the same model. The current model, VGG-X, is intended to work on the CIFAR-10 dataset: traditionally, the model works on the ImageNet \cite{ILSVRC15} dataset, but as we wanted to reduce \ndif{} by reusing the same data split, we did not succeed in time to provide fast and deterministic serving of the ImageNet dataset. Furthermore training on such a dataset would have taken considerably more time.

We adapted the original VGG model (\pytorch's implementation in Torchvision \cite{torchvision}), to work on 32x32 input images. While this does not change the results for our study, in the sense that comparison was made for the same model, which achieves acceptable good performance (around 70-74\%) and considering that comparisons should be made only between \buggy~and \corrected~versions, our comparison holds. Notwithstanding the validity on a theoretical side, the neural network should achieve better performance considering its size. We believe that the adaptation of the model to fit the 32x32 input did in fact cause some non-optimal configuration for the network. Notably, there is one aspect of the network that caused some potential non-determinism: the dropout layer. Two dropout layers with activation probability of 0.5 were in the network: this could have affected the determinism of the computation. This falls under the \ndif~of ``Model definition'' which is not studied in this paper.

\subsection{Build system}\label{sec:empirical:limitations:build}
In the context of this study, we built versions of \pytorch{} using a common build process and using the default parameters for build execution. As most software are distributed prebuilt through various channels, the upside from this process is that we reduce the possibility of introducing multiple build-related errors, as the recommended configurations reflect most widespread use of the framework. There are, of course, differences that can affect the final build artifacts from the ones we have built, notably because there might be some changes in the specific tooling used. Indeed, in the aim of reducing \cfvars, a single build system configuration was chosen to build all the versions with two variants on the \Python{} version. Each variant represents a \Docker~container with the dependencies used to build \pytorch{} according to their guidelines. Nonetheless, both containers use \CUDA~\API~\version{10.1}. Note that building does not use the GPU, but here the \CUDA~version refers to the \CUDA~toolkit, \ie~the necessary developer dependencies needed to build application with support for \CUDA.

The methodology prescribes using a build system in order to build \mversions~versions. However, great care must be taken when doing so: in the case of \pytorch{}, there are multiple computational ``backends''. These are library dependencies that exist in order to execute mathematical operations allowing different hardware and software configurations to run \pytorch{}. The limitation comes from the fact that some parts of the library can only be used, \ie compiled with one of these ``backends''. In an ideal world, one would need to build all the different versions of a framework with the different build options and then make an assessment of the impact of those while using statically linked libraries. Practically, this is infeasible and means that care has to be taken when building a version for a specific bug.

For reproducibility, the best way to build a framework such as \pytorch{} would be to statically link every dependency instead of using dynamic dependencies as this would mitigate improper or deviation of the methodology. Indeed, when statically linking an assembly, there would be no question of which runtime on the framework dependencies would be used. This would allow relaxation of the current constraint of running within the \Docker~container. Configuring static linking, however, is not trivial and would require a lot of effort. Furthermore, since we use \Docker~to control the runtime dependencies, we have tight constraints on the available runtime versions.

While it would have been possible to only use one \Python{} version for all the bugs, there is no guarantee that a specific version of \Python{} would be supported in a specific version of \pytorch{}. As an example, it is  possible to build an old version of \pytorch{}, ex  using \version{1.1.0} released in April 2019 with a \Python{} version \version{3.7.9} released in August 2020, this configuration would not have been possible at the release of said \pytorch{} version and there is no practical use to ensure compatibility retroactively.

\section{Conclusion and Future works}\label{sec:future-works}
In this paper, we first introduced \framework{}, a tool set to minimize the effects on non-determinism in ML experiments. It performs both random-seed control and data split control, a primitive form of data versioning. Then, we demonstrated an application of \framework{} to measure the impact of any ``change'' in a ML repository by using the commits as the revision upon which a build process would be initiated. We also explored the various factors that influence reproducibility in this context and devised the methodology and artifacts used in order to enhance reproducibility in the study. The aspect of reproducibility is showcased by (1) reducing the \cfvars~of the methodology and measurements by reducing the \ndif~using \framework{} and (2) by the reproducibility of this study. Finally, we conducted an empirical case study on \pytorch, as a popular \dl~framework, by extracting, filtering, building and evaluating the artifacts. We have collected \nBugsCollected~bugs from \pytorch's~repository, labelled \nBugsLabelled, filtered \nBugsUsedForStudy, built \nBugsBuilt~bugs (for their \buggy~and \corrected~version) and conducted our statistical analysis on \nBugsResults~of them. Based on these numbers and our dataset, we can not conclude that our sample of bugs from \pytorch~has an impact on model performance.

With the current tooling presented in this work and by following a similar methodology, several other empirical research on ML configurations can be made. A very first one is investigating a more comprehensive set of bugs in ML frameworks to evaluate their impact on ML application. Moreover, one may study how the dependencies' versions in ML frameworks can affect the performance metrics of ML models or how the use of deprecated APIs in ML frameworks can affect the performance metrics of ML models. Instead of building the versions which did contain a bug and its following revision, one may inject the changes in a stable version. This would partially solve the problems about possible missing versions, as the build process for a stable release has more chance to be robust and have longevity. We demonstrated an application of \framework{} by analyzing the effect of bugs but its applicability is not limited to such scope.  \framework{} can be utilized for various studies where controlled ML experiments are inevitable. For instance, our study on bugs can be extended to cover SGD convergence rate, while it is hard to run such a study without a tool like \framework{}. Moreover, one can extend \framework{} to cover the reproducibility of ML software in general that will include the presented regression tests in this paper.  
\bibliography{ref.bib}
\bibliographystyle{ieeetr}

\end{document}